\documentclass[preprint]{revtex4}
\begin{document}
\title{Nonextensive Entropies \\
derived from Form Invariance of Pseudoadditivity}
\author{Hiroki Suyari}
\affiliation{%
Department of Information and Image Sciences,
Faculty of Engineering, Chiba University\\
1-33, Yayoi-cho, Inage-ku, Chiba-shi, Chiba, 263-8522 Japan
}%
\email{suyari@tj.chiba-u.ac.jp}
\date{\today\\ \bigskip\bigskip\bigskip\bigskip\bigskip}

\begin{abstract}
The form invariance of pseudoadditivity
is shown to determine the structure of nonextensive entropies.
Nonextensive entropy is defined as the appropriate expectation value of
nonextensive information content, similar to the definition of Shannon entropy.
Information content in a nonextensive system is obtained uniquely
from generalized axioms by replacing the usual additivity with pseudoadditivity.
The satisfaction of the form invariance of the pseudoadditivity
of nonextensive entropy and its information content
is found to require the normalization of nonextensive entropies.
The proposed principle requires the same normalization as that derived
in [A.K. Rajagopal and S. Abe, Phys. Rev. Lett. {\bf 83}, 1711 (1999)],
but is simpler and establishes a basis
for the systematic definition of various entropies in nonextensive systems.
\end{abstract}
\maketitle

\section{Introduction}
Since the first proposal of nonextensive entropy by Tsallis \cite{Ts88,CT91},
there have been many successful studies and applications analyzing physical
systems such as long-range interactions, long-time memories, and multi-fractal
structures in the nonextensively generalized Boltzmann-Gibbs
statistical mechanics \cite{AO01}.
In the rapid progress in this field, 
some modifications have been made to mathematical formulations
in the generalized statistical mechanics
in order to maintain the physical consistency.
One of the most important modifications was the introduction of
an appropriate definition of the generalized expectation value.
This modification has already appeared in the literature
\cite{TMP98}, but has been applied
as a candidate for satisfying the physical requirements in a given situation
without a systematic framework.
The necessity to apply such modifications invites the establishment of guiding principles
that will provide a clear basis for proposed generalizations of 
Boltzmann-Gibbs statistical mechanics.
In a recent paper \cite{RA99}, Rajagopal and Abe
presented a principle for determining the structure of nonextensive entropies.
Their principle was the form invariance of the Kullback-Leibler
entropy when generalized to nonextensive situations.

In the present paper, a much simpler principle
for determining the structure of nonextensive entropies is presented.
The original Tsallis entropy is determined for the given appropriate axioms
\cite{Sa97,Ab00,Fa98}. 
In contrast to these axiomatic approaches, we define nonextensive entropy in another way;
in terms of the appropriate expectation value of nonextensive information content
similar to the definition for Shannon entropy \cite{VO79}.
This definition has already been applied to the generalization of the
Shannon source coding theorem using the {\it normalized $q$-expectation value}
of {\it nonextensive information content} \cite{Ya01}.
Nonextensive information content $I_q\left( p \right)$
is defined by $I_q\left( p \right)=-\ln _qp$ in \cite{Ya01}
as an intuitively natural generalization of the standard information content
$I_1\left( p \right)=-\ln p$ (referred to as {\it self-information}
or {\it self-entropy} in Shannon information theory \cite{VO79})),
where $\ln _qx$ is a $q$-logarithm function defined by 
$\ln _qx\equiv{{\left( {x^{1-q}-1} \right)} \mathord{\left/ {\vphantom {{\left(
{x^{1-q}-1} \right)} {\left( {1-q} \right)}}} \right. \kern-\nulldelimiterspace}
{\left( {1-q} \right)}}$.
However, in \cite{Ya01}, the form invariance presented in this paper was
not mentioned. 
We introduce the axioms for nonextensive information content
$I_q\left( p \right)$ as a slight generalization of that for the standard
information content, and obtain $I_q\left( p \right)$ uniquely 
from the generalized axioms. 
The requirement of form invariance of pseudoadditivity
when we define nonextensive entropy $S_q\left( p \right)$
as the appropriate expectation value of $I_q\left( p \right)$,
\begin{equation}
S_q\left( p \right)\equiv E_{q,p}\left[ {I_q\left( p_i \right)} \right],
\end{equation}
leads to the determination of the structure of the nonextensive entropy,
where $E_{q,p}\left[ {\;\cdot \;} \right]$ is the expectation value
satisfying the following form invariance of the pseudoadditivity:
\begin{equation}
{I_q\left( {p_1p_2} \right)\over{k}}
={I_q\left( {p_1} \right)\over{k}}
+{I_q\left( {p_2} \right)\over{k}}
+\varphi\left( q \right)\cdot
{I_q\left( {p_1} \right)\over{k}}\cdot{I_q\left( {p_2} \right)\over{k}}
\end{equation}
and
\begin{equation}
{S_q\left( {p^{AB}} \right)\over{k}}
={S_q\left( {p^A} \right)\over{k}}
+{S_q\left( {p^B} \right)\over{k}}+
\varphi\left( q \right)\cdot
{S_q\left( {p^A} \right)\over{k}}\cdot{S_q\left( {p^B} \right)\over{k}}
\end{equation}
where $k$ is a positive constant.
$\varphi\left( q \right)$ is any function of the nonextensivity
parameter $q$ and satisfies the conditions (\ref{boundary}) given below.

Note that information content means the amount of information
provided by a result of an observation in a physical sense.
The standard information content $I_1\left( p \right)=-\ln p$
has been called {\it surprise} by Watanabe \cite{Wa69},
and {\it unexpectedness} by Barlow \cite{Ba90}.

\section{Nonextensive Self-information}

The axioms of standard information content $I_1:\left[ {0,1} \right]\to R^+$ satisfying
$I_1\left( 1 \right)=0$ are given as follows
\cite{VO79}:

\begin{description}
\item{[S1]} $I_1$ is differentiable with respect
to any $p\in\left( {0,1}\right)$,
\item{[S2]} $I_1\left( {p_1p_2} \right)=I_1\left( {p_1}
\right)+I_1\left( {p_2} \right)$ for any $p_1,p_2\in \left[ {0,1} \right]$.
\end{description}

Axiom [S2] means that the information content for two stochastically
independent events is given by the sum of the two sets of information.

For the above axioms, $I_1\left( p \right)$ is determined uniquely by
\begin{equation}
I_1\left( p \right)=- k \ln p
\end{equation}
where $k$ is a positive constant \cite{VO79}.

The above axioms are generalized in nonextensive situations as follows.

Nonextensive information content 
$I_q:\left[ {0,1} \right]\to R^+$ for any fixed $q\in R^+$,
satisfying
\begin{equation}
\mathop {\lim }\limits_{q\to 1}I_q\left( p \right)=I_1\left( p \right)=- k \ln p,
\label{initial}
\end{equation}
should have the following properties.

\begin{description}
\item{[T1]} $I_q$ is differentiable
with respect to any $p\in\left( {0,1}\right)$ and $q\in R^+$,
\item{[T2]} $I_q\left( {p} \right)$
is convex with respect to $p\in \left[0,1\right]$ for any fixed $q\in R^+$,
\item{[T3]} there exists a function $\varphi :R\to R$ such that
\begin{equation}
{I_q\left( {p_1p_2} \right)\over{k}}={I_q\left( {p_1}\right)\over{k}}
+{I_q\left( {p_2} \right)\over{k}}
+\varphi\left( q \right)\cdot{I_q\left( {p_1}\right)\over{k}}\cdot
{I_q\left( {p_2} \right)\over{k}}
\label{pseudoadditivity}
\end{equation}
 for any $p_1,p_2\in \left[ {0,1} \right]$,
where $\varphi \left( q \right)$ is differentiable with respect to any $q\in R^+$,
\begin{equation}
\mathop {\lim }\limits_{q\to 1}{{d\varphi \left( q \right)} \over {dq}}\ne 0,\;\;
\mathop {\lim }\limits_{q\to 1}\varphi \left( q \right)=\varphi \left( 1 \right)=0,
\;\;\hbox{and}\;\;
\varphi \left( q \right)\ne 0\; \left( {q\ne 1} \right).
\label{boundary}
\end{equation}
\end{description}


Equation (\ref{pseudoadditivity}) is called
{\it pseudoadditivity} in many studies \cite{AO01}, as a special form of {\it
composability} \cite{HJ99,Ab01}.

Note that in these generalized axioms, [T2] is needed 
to maintain nonnegativity of the Kullback-Leibler entropy for any $q\in R^+$
when generalized to nonextensive situations \cite{Ta98,BPT98}.
In general, Kullback-Leibler entropy $K\left( {p^A\left\| {p^B} \right.} \right)$
is defined by the appropriate expectation value of the difference 
between two information contents
\cite{VO79}.
\begin{equation}
K\left( {p^A\left\| {p^B} \right.} \right)
\equiv E_{p^A}\left[ {I\left({p_i^B}\right)-I\left( {p_i^A} \right)} \right]
\label{KLentropy}
\end{equation}
Therefore the nonnegativity of the Kullback-Leibler entropy leads to
Gibbs inequality \cite{Gi48,Is71}:
\begin{equation}
K\left( {p^A\left\| {p^B} \right.} \right)\ge 0\quad \Leftrightarrow \quad 
S\left( p^A \right)=E_{p^A}\left[ I\left({p_i^A}\right)\right]\le
E_{p^A}\left[ I\left({p_i^B}\right)\right].
\end{equation}
When $E_{p}$ is a normalized expectation value (i.e., $E_{p}\left[1\right]=1$)
and $p^B=\left( {{1 \over W},\;\cdots ,\;{1 \over W}}\right)$,
the right-hand side $E_{p^A}\left[ I\left({p_i^B}\right)\right]$ of 
the above Gibbs inequality is equal to the maximum entropy.
\begin{equation}
E_{p^A}\left[ I\left({1 \over W}\right)\right]
=I\left( {{1 \over W}} \right)
=E_{{\raise3pt\hbox{$\scriptstyle 1$} \!\mathord{\left/ {\vphantom
{\scriptstyle {1 W}}}\right.\kern-\nulldelimiterspace} \!\lower3pt\hbox{$\scriptstyle
W$}}}\left[ {I\left( {{1 \over W}} \right)} \right]
=S\left( {{1 \over W},\;\cdots ,\;{1 \over
W}} \right)
\label{maximum entropy principle}
\end{equation}
This inequality coincides with the maximality condition
which is one of the Shannon-Khinchin axioms \cite{SW63,Kh57},
that is, the simplest form of the maximum entropy principle without constraints.
Therefore the satisfaction of the nonnegativity of the Kullback-Leibler entropy
for any $q\in R^+$ is needed when generalized to nonextensive systems.
In order to satisfy the requirement of the nonnegativity,
the information content $I_q\left( p \right)$
should be convex with respect to $p\in \left[0,1\right]$ for any fixed $q\in R^+$
and its corresponding appropriate expectation value should be applied.
Such examples of expectation value are {\it $q$-expectation value} 
and {\it normalized $q$-expectation value} \cite{CT91,TMP98}.
Under these two conditions (convex information content and appropriate expectation value),
the nonnegativity of the nonextensive Kullback-Leibler can be proved.
The proof is given in appendix A.


Using the axioms [T1]$\sim$[T3],
we determine $I_q\left( p \right)$ uniquely in the following
procedures.
Using (\ref{pseudoadditivity}), for any $1+\Delta \in \left( {0,1} \right)$, we have
\begin{equation}
{I_q\left( {p\left( {1+\Delta } \right)} \right)\over{k}}
={I_q\left( {p+\Delta p}\right)\over{k}}
={I_q\left( p \right)\over{k}}
+{I_q\left( {1+\Delta } \right)\over{k}}
+\varphi \left( q \right)\cdot
{I_q\left(p \right)\over{k}}\cdot{I_q\left( {1+\Delta } \right)\over{k}}.
\end{equation}
This can be rewritten as
\begin{equation}
{{I_q\left( {p+\Delta p} \right)-I_q\left( p \right)} \over {\Delta p}}
={1\over{k}}\cdot{{I_q\left( {1+\Delta } \right)}\over \Delta }\cdot{{k+\varphi
\left( q \right)I_q\left( p \right)} \over p}.
\label{differntial0}
\end{equation}
Taking the limit $\Delta \to 0$ on both sides of equation
(\ref{differntial0}), we obtain
\begin{equation}
{{dI_q\left( p \right)} \over {dp}}={\beta\over{k}}\cdot {{k+\varphi \left( q
\right)I_q\left( p \right)} \over p}
\end{equation}
where $\beta \equiv\mathop {\lim }\limits_{\Delta\to 0-}{{I_q\left( {1+\Delta }
\right)} \over \Delta }$
and the first axiom [T1] is applied.
The differential equation is given by
\begin{equation}
{1 \over {k+\varphi \left( q \right)y}}dy={\beta\over{k}}\cdot {1 \over x}dx
\end{equation}
where $x\equiv p$ and $y\equiv I_q\left( p \right)$.
This can be solved analytically;
the rigorous solution is
\begin{equation}
y=k\cdot{{\left( {Cx^\beta } \right)^{{\varphi \left( q \right)}\over{k}}-1} \over {\varphi
\left( q
\right)}},\quad \hbox{that is,} \quad
I_q\left( p \right)=k\cdot{{\left( {Cp^\beta } \right)^{{\varphi \left( q \right)}\over{k}}-1}
\over {\varphi \left( q \right)}}
\end{equation}
where $C$ is a constant.
Then, from the initial condition
$\mathop {\lim }\limits_{q\to 1}I_q\left( p \right)=I_1\left( p \right)=- k \ln p$,
we have that $C=1$ and $\beta =-k$, where
conditions (\ref{boundary}) are applied.
Thus, the nonextensive information content
$I_q\left( p \right)$ is derived as
$I_q\left( p \right)
=k\cdot{{p^{- \varphi \left( q \right)}-1} \over {\varphi \left( q\right)}}.$

Moreover, by [T2], the second differential of ${I_q\left( p \right)}$
with respect to $p$ should be nonnegative for any fixed $q\in R^+$.
Thus, we can derive a constraint $\varphi \left( q \right)+1\ge 0$ for any $q\in R^+$.

Summarizing these results, the nonextensive information content
$I_q\left( p \right)$ obtained from the axioms [T1]$\sim$[T3] is 
\begin{equation}
I_q\left( p \right)
=k\cdot{{p^{- \varphi \left( q \right)}-1} \over {\varphi \left( q\right)}}
\label{information content}
\end{equation}
where $k$ is a positive constant and
\begin{equation}
\varphi \left( q \right)+1\ge 0\quad  {\hbox{for any}}\;{q\in R^+}.
\label{varphi_condition}
\end{equation}
For example, $\varphi \left( q \right)=q-1$ implies $I_q\left( p \right)=-k\ln_qp$.

Note that there have been already remarks on the alternative candidates
for nonextensive information content
to define the original Tsallis entropy \cite{AO01}.
They are
\begin{equation}
I_q^{\left( 1 \right)}\left( p \right)\equiv -k\ln _qp
\quad\hbox{and}\quad
I_q^{\left( 2 \right)}\left( p \right)\equiv k\ln _qp^{-1},
\end{equation}
and they only coincide for $q=1$.
\begin{equation}
S_q^{\hbox{org}}\left( p \right)\equiv k\cdot{{1-\sum\limits_{i=1}^W {p_i^q}} \over {q-1}}
=\sum\limits_{i=1}^W {p_i^q I_q^{\left( 1 \right)}\left( p_i \right)}
=\sum\limits_{i=1}^W {p_i I_q^{\left( 2 \right)}\left( p_i \right)}
\label{2information contents}
\end{equation}
$I_q^{\left( 1 \right)}\left( p \right)$ and $I_q^{\left( 2 \right)}\left( p \right)$
correspond to 
$\varphi^{\left( 1 \right)} \left( q \right)\equiv q-1$ and 
$\varphi^{\left( 2 \right)} \left( q \right)\equiv 1-q$ as $\varphi \left( q \right)$
in (\ref{information content}), respectively.
However, the latter case $\varphi^{\left( 2 \right)} \left( q \right)=1-q$ does {\it not}
satisfy the identity (\ref{varphi_condition}) for any $q\in R^+$, that is,
$I_q^{\left( 2 \right)}\left( p \right)$ does {\it not} possess the property of {\it convexity}
[T2]. Therefore, $I_q^{\left( 2 \right)}\left( p \right)$ {\it cannot} be 
information content. 
Even if $I_q^{\left( 2 \right)}\left( p \right)$ is applied as information content,
then the nonnegativity property of
the nonextensive Kullback-Leibler entropy is {\it not} held for any $q\in R^+$ for lack of the
convexity of $I_q^{\left( 2 \right)}\left( p \right)$, as stated above.
The convexity of information content is applied to Jensen's inequality
in order to prove the nonnegativity of the nonextensive Kullback-Leibler entropy.
See the appendix A for the details.


In case of $\varphi \left( q \right)=q-1$ and $k=1$, the pseudoadditivity
(\ref{pseudoadditivity}) of
$I_q\left( p \right)$ is remarkably similar to the relation of 
the Jackson basic number in $q$-deformation theory \cite{Ab97,Jo98} as follows. 
Let $\left[ X\right]_q$ be the Jackson basic number of a physical quantity $X$, that is,
$\left[ X \right]_q\equiv{{\left( {q^X-1} \right)} \mathord{\left/ {\vphantom {{\left(
{q^A-1} \right)} {\left( {q-1} \right)}}} \right. \kern-\nulldelimiterspace} {\left(
{q-1} \right)}}$.
Then the Jackson basic number of $X+Y$ satisfies the identity
$\left[ {X+Y} \right]_q=\left[ X \right]_q+\left[ Y \right]_q+\left( {q-1}
\right)\left[ X \right]_q\left[ Y \right]_q$.
The surprising similarity to pseudoadditivity (\ref{pseudoadditivity})
can be seen if we consider a quantity 
$f\left( x \right)=p^{- \left({x-1} \right)}$.
Clearly, $f\left( 1 \right)=1$.
Standard information content $I_1\left( p \right)$ is expressed as
$I_1\left( p \right)=\left. {{{df\left( x \right)} \mathord{\left/ {\vphantom
{{df\left( x \right)} {dx}}} \right. \kern-\nulldelimiterspace} {dx}}} \right|_{x=1}$;
nonextensive information content
is given by $I_q\left( p \right)=D_q\left. {f\left( x \right)}
\right|_{x=1}\equiv\left. {{{\left( {f\left( {qx} \right)-f\left( x \right)} \right)}
\mathord{\left/ {\vphantom {{\left( {f\left( {qx} \right)-f\left( x \right)} \right)}
{\left( {qx-x} \right)}}} \right. \kern-\nulldelimiterspace} {\left( {qx-x} \right)}}}
\right|_{x=1}$, where $D_q$ is the Jackson differential.
According to $q$-deformation theory, the property 
$\mathop {\lim }_{q\to 1}I_q\left(p\right)=I_1\left( p \right)$
originates from the convergence
$\mathop {\lim }_{q\to 1}D_q={d \mathord{\left/ {\vphantom {d {dx}}} \right.
\kern-\nulldelimiterspace} {dx}}$.

\section{Effects of renormalization of nonextensive entropies}

The normalized nonextensive entropies follows
naturally from the form invariance between entropy 
and its information content.
In this section we assume $k=1$ for simplicity.

Similar to Shannon entropy, nonextensive entropy
$S_q\left( p \right)$ is defined as the expectation value of the
information content $I_q\left( p \right)$ obtained in (\ref{information content}).
For example, the nonextensive entropy $S_q^{\hbox{org}}$
using the unnormalized expectation value $E_{q,p}^{\hbox{org}}\left[ {\;\cdot \;} \right]$
is given by
\begin{equation}
S_q^{\hbox{org}}\left( p \right)= E_{q,p}^{\hbox{org}}\left[ {I_q\left( p \right)}
\right]\;\equiv\sum\limits_{i=1}^W {p_i^qI_q\left( {p_i} \right)}
\label{oldexpectation}
\end{equation}
where $W$ is the total number of microscopic configurations with probabilities
$\left\{ {p_i} \right\}$.
In the definition of $E_{q,p}^{\hbox{org}}$ in (\ref{oldexpectation}),
the {\it$q$-expectation value} \cite{CT91,TMP98} is used.
If we let
$\varphi \left( q \right)=q-1$, then 
$S_q^{\hbox{org}}\left( p \right)$ is concretely derived
from (\ref{information content}) and
(\ref{oldexpectation}) as follows:
\begin{equation}
S_q^{\hbox{org}}\left( p \right)={{1-\sum\limits_{i=1}^W {p_i^q}} \over {q-1}}.
\label{originalTsallis}
\end{equation}
This is the original Tsallis entropy \cite{Ts88}.

Let $A$ and $B$ be two independent systems in the sense that
the probabilities $p_{ij}^{AB}$ of the total system $A+B$
factorize into those of $A$ and of $B$, i.e.,
\begin{equation}
p_{ij}^{AB}=p_i^Ap_j^B\quad  {\hbox{for any}}\;\;{i=1,\;\cdots ,\;W_A}
\hbox{ and } {j=1,\;\cdots ,\;W_B}.
\label{factor}
\end{equation}
The nonextensive entropy $S_q^{\hbox{org}}\left( {p^{AB}} \right)$
of the total system $A+B$ can then be expanded using definitions
(\ref{oldexpectation}) and (\ref{factor}) as follows:
\begin{equation}
S_q^{\hbox{org}}\left( {p^{AB}} \right)=E_{q,p}^{\hbox{org}}\left[ {I_q\left( {p^{AB}}
\right)}
\right]=\sum\limits_{i=1}^{W_A} {\sum\limits_{j=1}^{W_B} {\left( {p_{ij}^{AB}}
\right)^qI_q\left( {p_{ij}^{AB}} \right)}}=\sum\limits_{i=1}^{W_A}
{\sum\limits_{j=1}^{W_B} {\left( {p_i^Ap_j^B} \right)^qI_q\left( {p_i^Ap_j^B}
\right)}}
\label{expansion}
\end{equation}
Applying the pseudoadditivity (\ref{pseudoadditivity}) for information content
$I_q\left( {p}\right)$, we obtain
\begin{eqnarray}
S_q^{\hbox{org}}\left( {p^{AB}} \right)&=&\sum\limits_{i=1}^{W_A}
{\sum\limits_{j=1}^{W_B} {\left( {p_i^Ap_j^B} \right)^q\left\{ {I_q\left( {p_i^A}
\right)+I_q\left( {p_j^B}
\right)+\varphi \left( q \right)I_q\left( {p_i^A} \right)I_q\left( {p_j^B} \right)}
\right\}}}\nonumber\\
&=&\left( {\sum\limits_{j=1}^{W_B} {\left( {p_j^B} \right)^q}} \right)S\left( {p^{A}}
\right)+\left( {\sum\limits_{i=1}^{W_A} {\left( {p_i^A} \right)^q}} \right)
S\left({p^{B}}\right)+\varphi \left( q \right)S\left( {p^{A}} \right)S\left( {p^{B}}
\right)
\label{expand1}
\end{eqnarray}
where we used
\begin{equation}
S^{\hbox{org}}\left( {p^{A}} \right)=\sum\limits_{i=1}^{W_A} {\left( {p_i^A}
\right)^qI_q\left( {p_i^A}
\right)}\quad\hbox{and}\quad S^{\hbox{org}}\left( {p^{B}} \right)=\sum\limits_{j=1}^{W_B}
{\left( {p_j^B}
\right)^qI_q\left( {p_j^B} \right)}.
\end{equation}
Dividing both sides of (\ref{expand1}) by
\begin{equation}
\sum\limits_{i=1}^{W_A} {\sum\limits_{j=1}^{W_B} {\left(
{p_{ij}^{AB}} \right)^q}}=\left( {\sum\limits_{i=1}^{W_A} {\left( {p_i^A} \right)^q}}
\right)\!\!\left( {\sum\limits_{j=1}^{W_B} {\left( {p_j^B} \right)^q}} \right)
\left(\ne 0 \right)
\end{equation}
yields
\begin{equation}
{{S_q^{\hbox{org}}\left( {p^{AB}} \right)} \over 
{\sum\limits_{i=1}^{W_A} {\sum\limits_{j=1}^{W_B} {\left( {p_{ij}^{AB}}
\right)^q}}}}={{S_q^{\hbox{org}}\left( {p^A} \right)} \over {\sum\limits_{i=1}^{W_A}
{\left( {p_i^A} \right)^q}}}+{{S_q^{\hbox{org}}\left( {p^B} \right)} \over
{\sum\limits_{j=1}^{W_B} {\left( {p_j^B} \right)^q}}}+\varphi \left( q
\right){{S_q^{\hbox{org}}\left( {p^A} \right)} \over {\sum\limits_{i=1}^{W_A} {\left(
{p_i^A}
\right)^q}}} {{S_q^{\hbox{org}}\left( {p^B} \right)}
\over {\sum\limits_{j=1}^{W_B} {\left( {p_j^B} \right)^q}}}.
\label{obtainedrelation}
\end{equation}
In order to preserve the form of pseudoadditivity 
between nonextensive entropy $S_q\left( p \right)$
and the corresponding information content $I_q\left( {p} \right)$,
the nonextensive entropy is modified as
\begin{equation}
S_q^{\hbox{nor}}\left( p \right)\;\equiv{{S_q^{\hbox{org}}\left( p \right)} \over
{\sum\limits_{j=1}^W {{p_j^q}}}}
={{\sum\limits_{i=1}^W {p_i^qI_q\left( {p_i} \right)}}\over{\sum\limits_{j=1}^W {{p_j^q}}}}
\label{normalization0}
\end{equation}
which is the expectation value of the information content $I_q\left( {p_i} \right)$
with respect to the {\it escort distribution} \cite{BS93} of $p$ of order $q$.
Then, by substituting (\ref{normalization0}) into (\ref{obtainedrelation}),
the following equation with respect to
$S_q^{\hbox{nor}}\left( p \right)$ is obtained.
\begin{equation}
S_q^{\hbox{nor}}\left( {p^{AB}} \right)
=S_q^{\hbox{nor}}\left( {p^A}\right)
+S_q^{\hbox{nor}}\left( {p^B}\right)+\varphi \left( q \right)
S_q^{\hbox{nor}}\left( {p^A} \right) S_q^{\hbox{nor}}\left( {p^B} \right).
\label{modified_pseudoadditivity}
\end{equation}
This is the same pseudoadditivity of nonextensive entropy
$S_q\left( p \right)$ as (\ref{pseudoadditivity}).
We have thus derived
from the expectation value of $I_q\left( {p} \right)$
the form invariance of the pseudoadditivity
of nonextensive entropy and its information content.
Moreover, the nonextensive entropy $S_q^{\hbox{nor}}\left(p\right)$
defined by (\ref{normalization0}) 
is actually the {\it $q$-normalized nonextensive entropy}
\cite{LV98,RA99}.
Thus, according to the principle of the form invariance of pseudoadditivity
between nonextensive entropy $S_q\left( p \right)$
and its information content $I_q\left( {p} \right)$,
$S_q^{\hbox{nor}}\left( p \right)$
should be used as nonextensive entropy
instead of $S_q^{\hbox{org}}\left( p \right)$.
If $E_{q,p}^{\hbox{nor}}$ denotes the normalized $q$-expectation value
with respect to $\left\{ {p_i} \right\}$, defined by
\begin{equation}
E_{q,p}^{\hbox{nor}}\left[ A \right]\;
\equiv{{E_{q,p}^{\hbox{org}}\left[ A \right]} \over {\sum\limits_{j=1}^{W} {p_j^q}}}
={{\sum\limits_{i=1}^{W}{{p_i^q}A_i}} \over 
{\sum\limits_{j=1}^{W} {p_j^q}}}
\label{newexpectation}
\end{equation}
where $A$ is a physical quantity, then $q$-normalized nonextensive entropy
$S_q^{\hbox{nor}}\left( p \right)$ is given by
\begin{equation}
S_q^{\hbox{nor}}\left( p \right)
=E_{q,p}^{\hbox{nor}}\left[ I_q\left( {p_i}\right) \right].
\label{normalization1}
\end{equation}
For the most typical case 
$\varphi \left( q \right)=q-1$, $S_q^{\hbox{nor}}\left( p\right)$
is concretely given by
\begin{equation}
S_q^{\hbox{nor}}\left( p \right)
={{1-\sum\limits_{i=1}^W {p_i^q}} \over {\left(q-1\right){\sum\limits_{j=1}^W {{p_j^q}}}}}.
\label{modifiedTsallis}
\end{equation}
This normalized Tsallis entropy (\ref{modifiedTsallis})
is concave only if the nonextensivity parameter $q$
lies in $\left( {0,1} \right)$ \cite{AO01,RA99}.

Note that definition (\ref{oldexpectation}) can be easily
replaced with a more general unnormalized expectation value,
leading to almost the same conclusion as that derived in this study.
For example, if we use a more general form (\ref{information content})
as information content, then the expectation value
$E_{q,p}^{\hbox{g-org}}$ defined by (\ref{qexpectation})
can be applied to the definition of the generalized original Tsallis entropy
$S_q^{\hbox{g-org}}$.
\begin{equation}
S_q^{\hbox{g-org}}\left( p \right)
\equiv E_{q,p}^{\hbox{g-org}}\left[ {I_q\left( {p_i} \right)} \right]
=\sum\limits_{i=1}^W {p_i^{\varphi \left( q \right)+1}}I_q\left( p_i \right)
={{1-\sum\limits_{i=1}^W {p_i^{\varphi \left( q\right)+1}}}
\over  {\varphi \left( q\right)}}.
\label{g-originalTsallis}
\end{equation}
Then, along the same procedure as that presented in this section,
the following $S_q^{\hbox{g-nor}}$ can be obtained
in order to preserve the form invariance of pseudoadditivity
between nonextensive entropy and its information content.
\begin{equation}
S_q^{\hbox{g-nor}}\left( p \right)
\equiv{{S_q^{\hbox{g-org}}\left( p \right)} \over
{\sum\limits_{j=1}^W {{p_j^{\varphi \left( q \right)+1}}}}}
={{1-\sum\limits_{i=1}^W {p_i^{\varphi \left( q\right)+1}}}
\over  {\varphi \left( q\right){\sum\limits_{j=1}^W {{p_j^{\varphi \left( q \right)+1}}}}}}.
\label{g-normalization}
\end{equation}
In fact, when $\varphi \left( q\right)=q-1$,
the formulas (\ref{g-originalTsallis}) and (\ref{g-normalization})
coincide with (\ref{originalTsallis}) and (\ref{modifiedTsallis}), respectively.

A dissatisfaction of the form invariance of the pseudoadditivity
in the original Tsallis entropy can be revealed through the following simple calculation.
Here we take $k=1$ for simplicity.
When $\varphi \left( q \right)=q-1$,
and substituting (\ref{information content}) into (\ref{oldexpectation}),
$S_q^{\hbox{org}}\left( p \right)$ coincides with the original Tsallis entropy
\cite{Ts88} as shown in (\ref{originalTsallis}).
The original Tsallis entropy $S_q^{\hbox{org}}\left( p \right)$
given by (\ref{originalTsallis})
is widely known to satisfy the following pseudoadditivity \cite{AO01}:
\begin{equation}
S_q^{\hbox{org}}\left( {p^{AB}} \right)
=S_q^{\hbox{org}}\left( {p^A}\right)
+S_q^{\hbox{org}}\left( {p^B}\right)+ \left( 1-q \right)
S_q^{\hbox{org}}\left( {p^A} \right)S_q^{\hbox{org}}\left( {p^B} \right)
\label{entropy_pseudoadditivity}
\end{equation}
However, the pseudoadditivity (\ref{pseudoadditivity})
of $I_q\left( {p} \right)$
for the same condition (i.e., $\varphi \left( q \right)=q-1$) is
given by
\begin{equation}
I_q\left( {p_1p_2} \right)=I_q\left( {p_1}
\right)+I_q\left( {p_2} \right)+ \left( q-1 \right)I_q\left( {p_1}
\right)I_q\left( {p_2} \right).
\label{selfinformation_pseudoadditivity}
\end{equation}
By comparing (\ref{entropy_pseudoadditivity}) and 
(\ref{selfinformation_pseudoadditivity}),
the coefficient $(q-1)$ of the cross term
of pseudoadditivity (\ref{selfinformation_pseudoadditivity})
{\it differs} from the $(1-q)$ in (\ref{entropy_pseudoadditivity})
when $E_{q,p}^{\hbox{org}}\left[ {\;\cdot \;} \right]$
defined by (\ref{oldexpectation}) is used.
This clearly reveals that
the form of the pseudoadditivity of
$S_q^{\hbox{org}}\left( {p}\right)$ and $I_q\left( {p} \right)$
is {\it not}\, invariant
in the computation of $E_{q,p}^{\hbox{org}}\left[ {\;\cdot \;} \right]$.
In other words, the form of the pseudoadditivity is {\it not} fixed
when the unnormalized expectation value $E_{q,p}^{\hbox{org}}\left[ {\;\cdot \;} \right]$ is
applied to the definition of Tsallis entropy.

More generally, for the generalized original Tsallis entropy
$S_q^{\hbox{g-org}}$ obtained in (\ref{g-originalTsallis}),
the following pseudoadditivity is held.
\begin{equation}
S_q^{\hbox{g-org}}\left( {p^{AB}} \right)
=S_q^{\hbox{g-org}}\left( {p^A}\right)
+S_q^{\hbox{g-org}}\left( {p^B}\right)-{\varphi \left( q\right)}\cdot
S_q^{\hbox{g-org}}\left( {p^A} \right)S_q^{\hbox{g-org}}\left( {p^B} \right)
\label{g_entropy_pseudoadditivity}
\end{equation}
By comparing (\ref{pseudoadditivity}) and (\ref{g_entropy_pseudoadditivity}),
a \lq\lq${\varphi \left( q\right)}$ versus $-{\varphi \left( q\right)}$ inconsistency"
can be found as similar as the above discussion.
Note that when $q=1$, the form invariance discussed here holds
because both $E_{q,p}^{\hbox{org}}\left[ {\;\cdot \;} \right]$
and $E_{q,p}^{\hbox{g-org}}\left[ {\;\cdot \;} \right]$ become
a normalized expectation value when $q=1$.

Therefore the unnormalized expectation value such as $E_{q,p}^{\hbox{org}}\left[ {\;\cdot \;}
\right]$ result in an inconsistency in the form invariance of the pseudoadditivity
for the original Tsallis entropy.

If we let $\varphi \left( q \right)=q-1$, then from (\ref{modified_pseudoadditivity})
the following pseudoadditivity holds.
\begin{equation}
S_q^{\hbox{nor}}\left( {p^{AB}} \right)
=S_q^{\hbox{nor}}\left( {p^A}\right)
+S_q^{\hbox{nor}}\left( {p^B}\right)+ \left( q-1 \right)
S_q^{\hbox{nor}}\left( {p^A} \right)S_q^{\hbox{nor}}\left( {p^B} \right)
\label{entropy_pseudoadditivity2}
\end{equation}
In other words, $S_q^{\hbox{nor}}\left( p \right)$ given by (\ref{modifiedTsallis})
satisfies the same pseudoadditivity (\ref{entropy_pseudoadditivity2})
as (\ref{selfinformation_pseudoadditivity}).
Therefore, the form invariance of pseudoadditivity requires the change from
the familiar identity (\ref{entropy_pseudoadditivity}) to 
the modified one (\ref{entropy_pseudoadditivity2}).
This follows clearly from the above discussion, because when
$E_{q,p}^{\hbox{org}}\left[ {\;\cdot\;}\right]$ defined by (\ref{oldexpectation})
is applied to the definition of $S_q\left( p \right)$,
the form invariance of the pseudoadditivity is {\it not} held
as shown above.

Note that the obtained pseudoadditivity (\ref{entropy_pseudoadditivity2})
is same as the relation of the Jackson basic number:
$\left[ {X+Y} \right]_q=\left[ X \right]_q+\left[ Y \right]_q+\left( {q-1}
\right)\left[ X \right]_q\left[ Y \right]_q$
where $\left[ X \right]_q\equiv{{\left( {q^X-1} \right)} \mathord{\left/ {\vphantom {{\left(
{q^A-1} \right)} {\left( {q-1} \right)}}} \right. \kern-\nulldelimiterspace} {\left(
{q-1} \right)}}$ \cite{RA99}.
Consider a quantity 
${\tilde f}
\left( x \right)\equiv {1 \mathord{\left/ {\vphantom {1 {\sum\limits_i {\left( {p_i}
\right)^x}}}} \right. \kern-\nulldelimiterspace} {\sum\limits_i {\left( {p_i} \right)^x}}}$.
Clearly, ${\tilde f}\left( 1 \right)=1$.
Shannon entropy $S_1\left( p \right)$ is expressed as
$S_1\left( p \right)=\left. {{{d{\tilde f}\left( x \right)} \mathord{\left/ {\vphantom
{{d{\tilde f}\left( x \right)} {dx}}} \right. \kern-\nulldelimiterspace} {dx}}} \right|_{x=1}$;
normalized Tsallis entropy
is given by $S_q^{\hbox{nor}}\left( p \right)=D_q\left. {{\tilde f}\left( x \right)}
\right|_{x=1}\equiv\left. {{{\left( {{\tilde f}\left( {qx} \right)-{\tilde f}\left( x \right)}
\right)}
\mathord{\left/ {\vphantom {{\left( {{\tilde f}\left( {qx} \right)-{\tilde f}\left( x \right)}
\right)} {\left( {qx-x} \right)}}} \right. \kern-\nulldelimiterspace} {\left( {qx-x} \right)}}}
\right|_{x=1}$, where $D_q$ is the Jackson differential.

\section{Conclusion}
We have established a self-consistent principle
for the form invariance of the pseudoadditivity
of nonextensive entropy and its information content.
The present principle is drawn from Shannon information theory
and leads to the same normalization of the original Tsallis entropy
as that derived in \cite{RA99}.

Once a set of an information content and an expectation value is given,
various entropies such as Kullback-Leibler entropy (relative entropy) and mutual entropy
can be formulated systematically.
In nonextensive systems,
an information content (\ref{information content})
and two expectation values (\ref{oldexpectation}) and (\ref{newexpectation})
are given in the previous sections.
Therefore we can formulate two sets of various entropies based on two sets of
the information content (\ref{information content}) and 
the expectation value (\ref{oldexpectation}) or (\ref{newexpectation}), respectively.
Please see the concrete formulas in appendix B.

Note that the alternative selection from the original Tsallis entropy
or the normalized Tsallis entropy should be careful in each application.
From the mathematical point of view,
the normalized Tsallis entropy has nice properties
such as the form invariance of the pseudoadditivity,
the unified application of the normalized $q$-expectation value, 
the form invariance of the statement of the maximum entropy principle and so on.
However, from the physical point of view,
the original Tsallis entropy has many advantages over the normalized version.
For example, the results derived from the Kolmogorov-Sinai entropy
using the original Tsallis entropy have the perfect matching
with nonlinear dynamical behavior
such as the sensitivity to the initial conditions in chaos.
On the other hand, the normalized Tsallis entropy does not have these convenient properties.
Please see the references \cite{TPZ97,CLPT97,LT98,BTAO02} for the details.


The principle discussed here is based on the usual formulation
for information in Shannon information theory.
Therefore, the ideas presented in this paper are
an application of information theory to statistical mechanics,
similar to the philosophy of Jaynes' work \cite{Ja83}.
There remain many other applications of Shannon information theory
to this interesting field.

\begin{acknowledgments}
The author would like to thank Prof. Yoshinori Uesaka and Prof. Makoto Tsukada 
for their valuable comments and discussions.
The author is grateful to Prof. Sumiyoshi Abe for reading the first draft
and his useful comments.
The author is very much obliged to the second referee for his significant comments. 

\end{acknowledgments}


\appendix

\section{Gibbs inequality derived from convexity of information content
and appropriate expectation value}

As presented in (\ref{KLentropy}),
the Kullback-Leibler entropy is generally defined by means of the information content,
\begin{equation}
K_q\left( {p^A\;\left\| {\;p^B} \right.} \right)
=E_{q,p^A}\left[ {I_q\left(
{p_i^B} \right)-I_q\left( {p_i^A} \right)} \right].
\label{KLentropy1}
\end{equation}
Our result (\ref{information content}) implies that
\begin{equation}
{I_q\left( {p_2} \right)-I_q\left( {p_1} \right)
=p_1^{-\varphi \left( q \right)}I_q\left(
{{{p_2} \over {p_1}}} \right)}.
\end{equation}
Substituting this relation into (\ref{KLentropy1}),
the Kullback-Leibler entropy is
\begin{equation}
K_q\left( {p^A\;\left\| {\;p^B} \right.} \right)
=E_{q,p^A}\left[ {\left( {p_i^A} \right)^{-\varphi \left( q \right)}I_q\left( {{{p_i^B}
\over {p_i^A}}} \right)} \right].
\end{equation}
If we take an unnormalized expectation value:
\begin{equation}
E_{q,p}^{\hbox{g-org}}\left[ X \right]\equiv\sum\limits_{i=1}^W {p_i^{\varphi \left( q
\right)+1}}X_i,
\label{qexpectation}
\end{equation}
then
\begin{equation}
K_q\left( {p^A\;\left\| {\;p^B} \right.} \right)
=\sum\limits_{i=1}^W {p_i^AI_q\left( {{{p_i^B} \over {p_i^A}}} \right)}.
\label{KLentropy2}
\end{equation}
$I_q\left(p \right)$ is a convex function with respect to 
$p\in \left[0,1\right]$ for any $q\in R^+$.
Thus, we can use Jensen's inequality \cite{VO79}: if $f$ is a
convex function and $X$ is a physical quantity (random variable in mathematics), then
\begin{equation}
E\left[ {f\left( X \right)} \right]\ge f\left( {E\left[ X \right]} \right)
\end{equation}
where $E$ is the usual expectation value when $q=1$.
Therefore (\ref{KLentropy2}) satisfies
\begin{equation}
K_q\left( {p^A\;\left\| {\;p^B} \right.} \right)
=\sum\limits_{i=1}^n {p_i^AI_q\left( {{{p_i^B} \over {p_i^A}}} \right)}
\ge I_q\left( {\sum\limits_{i=1}^W {p_i^A{{p_i^B} \over {p_i^A}}}} \right)
=I_q\left( {\sum\limits_{i=1}^W {p_i^B}} \right)=I_q\left( 1 \right)=0.
\label{Giibsinequality}
\end{equation}
If we take a normalized expectation value:
\begin{equation}
E_{q,p}^{\hbox{g-nor}}\left[ X \right]\equiv\sum\limits_{i=1}^W {{{p_i^{\varphi \left( q
\right)+1}}
\over {\sum\limits_{j=1}^n {p_j^{\varphi \left( q \right)+1}}}}}X_i,
\label{noramlizedqexpectation}
\end{equation}
then the same Gibbs inequality as (\ref{Giibsinequality}) can be obtained.
When $\varphi \left( q \right)=q-1$, the expectation values
(\ref{qexpectation}) and (\ref{noramlizedqexpectation})
coincide with {\it $q$-expectation value} defined by (\ref{oldexpectation})
and {\it normalized $q$-expectation value} defined by (\ref{newexpectation}),
respectively.
Thus, if an expectation value is chosen appropriately for a given self-information,
then the nonnegativity condition of the Kullback-Leibler entropy is held.

\section{Systematic formulations of various nonextensive entropies}

The proposed procedure for defining nonextensive entropy
using an information content and an expectation value
is applicable to the systematic formulations of various nonextensive entropies
such as Kullback-Leibler entropy (relative entropy) and mutual entropy
in accordance with the formulations in Shannon information theory \cite{VO79}.

In nonextensive systems,
an information content and two expectations are given in section III.
Therefore we can formulate various nonextensive entropies
in the following two cases.
\begin{description}
\item{Case 1 :} the information content (\ref{information content}) and 
the $q$-expectation value (\ref{oldexpectation})
\item{Case 2 :} the information content (\ref{information content}) and 
the normalized $q$-expectation value (\ref{newexpectation})
\end{description}

In each formulation $\varphi \left( q \right)=q-1$ is used
as the most typical function of $\varphi \left( q \right)$.

\subsection{Various nonextensive entropies using $q$-expectation value
$E_{q}^{\hbox{org}}\left[ \cdot\right]$}

The information content (\ref{information content})
and the $q$-expectation value (\ref{oldexpectation}) are applied to
the definitions of various entropies as follows.

\begin{description}
\item{(O-1)} (nonextensive joint entropy) 
\begin{equation}
S_q^{\hbox{org}}\left( {p^{AB}} \right)\equiv E_{q,p^{AB}}^{\hbox{org}}\left[
{I_q\left( {p_{ij}^{AB}} \right)} \right]
={{1-\sum\limits_{i=1}^{W_A} {\sum\limits_{j=1}^{W_B} {\left(
{p_{ij}^{AB}} \right)^q}}} \over { {q-1} }},
\end{equation}
\item{(O-2)} (nonextensive conditional entropy) 
\begin{eqnarray}
S_q^{\hbox{org}}\left( {p^{B\left| A \right.}} \right)&\equiv& E_{q,p^A}^{\hbox{org}}\left[
{S_q\left( {{{p_{i*}^{AB}} \over {p_i^A}}} \right)} \right]
=\sum\limits_{i=1}^{W_A} {\left( {p_i^A} \right)^q
\left[ {{{1-\sum\limits_{j=1}^{W_B} {\left(
{{{p_{ij}^{AB}} \over {p_i^A}}} \right)^q}} \over {q-1}}} \right]}\nonumber\\
&=&{{\sum\limits_{i=1}^{W_A}
{\left( {p_i^A} \right)^q}-\sum\limits_{i=1}^{W_A} {\sum\limits_{j=1}^{W_B} {\left(
{p_{ij}^{AB}} \right)^q}}} \over {q-1}}
\end{eqnarray}
where $\left\{ {{{p_{i*}^{AB}} \over {p_i^A}}} \right\}=\left\{ {{{p_{i1}^{AB}} \over
{p_i^A}},\;\cdots ,\;{{p_{iW_B}^{AB}} \over {p_i^A}}} \right\}$
for $i=1,\;\cdots ,\;W_A$,
\item{(O-3)} (nonextensive Kullback-Leibler entropy)
\begin{equation}
K_q^{\hbox{org}}\left( {p^A\left\| {p^B} \right.} \right)
\equiv E_{q,p^A}^{\hbox{org}}\left[ {I_q\left({p_i^B}
\right)-I_q\left( {p_i^A} \right)} \right]
={{1-\sum\limits_{i=1}^W {p_i^B\left( {{{p_i^A} \over {p_i^B}}} \right)^q}} \over { {1-q}
}},
\label{definitionKLA}
\end{equation}
\item{(O-4)} (nonextensive mutual entropy)
\begin{eqnarray}
{\mathcal I}_q^{\hbox{org}}\left( {p^A;p^B} \right)&\equiv&
K_q^{\hbox{org}}\left( {p^{AB}\left\| {p^Ap^B} \right.}\right)
= E_{q,p^{AB}}^{\hbox{org}}
\left[ {I_q\left( {p_i^Ap_j^B} \right)-I_q\left( {p_{ij}^{AB}} \right)}
\right]\nonumber\\
&=&{{1-\sum\limits_{i=1}^{W_A} {\sum\limits_{j=1}^{W_B} {\left( {p_i^Ap_j^B} \right)\left(
{{{p_{ij}^{AB}} \over {p_i^Ap_j^B}}} \right)^q}}} \over { {1-q}}}.
\label{mutual_entropyA}
\end{eqnarray}
\end{description}
%

\subsection{Various nonextensive entropies using normalized $q$-expectation value
$E_{q}^{\hbox{nor}}\left[ \cdot\right]$}

The information content (\ref{information content})
and the $q$-expectation value (\ref{newexpectation}) are applied to
the definitions of various entropies as follows.

\begin{description}
\item{(N-1)} (nonextensive joint entropy) 
\begin{equation}
S_q^{\hbox{nor}}\left( {p^{AB}} \right)\equiv E_{q,p^{AB}}^{\hbox{nor}}\left[
{I_q\left( {p_{ij}^{AB}} \right)} \right]
={{1-\sum\limits_{i=1}^{W_A} {\sum\limits_{j=1}^{W_B} {\left(
{p_{ij}^{AB}} \right)^q}}} \over {\left( {q-1} \right)\sum\limits_{i=1}^{W_A}
{\sum\limits_{j=1}^{W_B} {\left( {p_{ij}^{AB}} \right)^q}}}},
\end{equation}
\item{(N-2)} (nonextensive conditional entropy) 
\begin{equation}
S_q^{\hbox{nor}}\left( {p^{B\left| A \right.}} \right)\equiv E_{q,p^A}^{\hbox{nor}}\left[
{S_q\left( {{{p_{i*}^{AB}} \over {p_i^A}}} \right)} \right]={{\sum\limits_{i=1}^{W_A} {\left[
{\left( {p_i^A} \right)^q{{1-\sum\limits_{s=1}^{W_B} {\left( {{{p_{is}^{AB}} \over {p_i^A}}}
\right)^q}} \over {\left( {q-1} \right)\sum\limits_{t=1}^{W_B} {\left( {{{p_{it}^{AB}} \over
{p_i^A}}} \right)^q}}}} \right]}} \over {\sum\limits_{j=1}^{W_A} {\left( {p_j^A} \right)^q}}}
\end{equation}
where $\left\{ {{{p_{i*}^{AB}} \over {p_i^A}}} \right\}=\left\{ {{{p_{i1}^{AB}} \over
{p_i^A}},\;\cdots ,\;{{p_{iW_B}^{AB}} \over {p_i^A}}} \right\}$
for $i=1,\;\cdots ,\;W_A$,
\item{(N-3)} (nonextensive Kullback-Leibler entropy)
\begin{equation}
K_q^{\hbox{nor}}\left( {p^A\left\| {p^B} \right.} \right)
\equiv E_{q,p^A}^{\hbox{nor}}\left[ {I_q\left({p_i^B}
\right)-I_q\left( {p_i^A} \right)} \right]
={{1-\sum\limits_{i=1}^W {p_i^B\left( {{{p_i^A} \over {p_i^B}}} \right)^q}} \over {\left( {1-q}
\right)\sum\limits_{j=1}^W {\left( {p_j^A} \right)^q}}},
\label{definitionKLB}
\end{equation}
\item{(N-4)} (nonextensive mutual entropy)
\begin{eqnarray}
{\mathcal I}_q^{\hbox{nor}}\left( {p^A;p^B} \right)&\equiv&
K_q^{\hbox{nor}}\left( {p^{AB}\left\| {p^Ap^B} \right.}\right)
= E_{q,p^{AB}}^{\hbox{nor}}
\left[ {I_q\left( {p_i^Ap_j^B} \right)-I_q\left( {p_{ij}^{AB}} \right)}
\right]\nonumber\\
&=&{{1-\sum\limits_{i=1}^{W_A} {\sum\limits_{j=1}^{W_B} {\left( {p_i^Ap_j^B} \right)\left(
{{{p_{ij}^{AB}} \over {p_i^Ap_j^B}}} \right)^q}}} \over {\left( {1-q}
\right)\sum\limits_{s=1}^{W_A} {\sum\limits_{t=1}^{W_B} {\left( {p_{st}^{AB}} \right)^q}}}}.
\label{mutual_entropyB}
\end{eqnarray}
\end{description}

Note that both of nonextensive mutual entropies respectively defined by (\ref{mutual_entropyA})
and (\ref{mutual_entropyB}) are clearly {\it symmetric} with respect to
\lq\lq$\left\{ {p_i^A} \right\}\leftrightarrow \left\{ {p_j^B}\right\}$"
in each formulation.
Furthermore, the {\it nonnegativity} of mutual entropies (\ref{mutual_entropyA})
and (\ref{mutual_entropyB}) 
is directly derived from that of the nonextensive Kullback-Leibler entropy, i.e.,
\begin{equation}
{\mathcal I}_q\left( {p^A;p^B} \right)= 
K_q\left( {p^{AB}\left\| {p^Ap^B} \right.}\right)\ge 0
\quad  {\hbox{for any}}\;{q\in R^+},
\end{equation}
with equality $p_{ij}^{AB}=p_i^Ap_j^B$ for any $i=1,\cdots ,W_A$
and $j=1,\cdots ,W_B$.


\end{document}